# On the role of selective nucleation and growth to recrystallization texture development in a Mg-Gd-Zn alloy

F. Mouhib[1*], B. Gao[2], T. Al-Samman[1]

1. Institute of Physical Metallurgy and Materials Physics, RWTH Aachen University, 52056 Aachen, Germany

2. Steel Institute (IEHK), RWTH Aachen University, 52056 Aachen, Germany

[*]Corresponding author, email mouhib@imm.rwth-aachen.de

**Abstract:**

One of the main material properties altered by rare earth additions in magnesium alloys is texture, which can be specifically adjusted to enhance ductility and formability. The current study aims at illuminating the texture selection process in a Mg-0.073at%Gd-0.165at%Zn alloy by investigating recrystallization nucleation and early nucleus growth during static recrystallization. An as-cast sample of the investigated alloy was deformed in uniaxial compression at 200°C till 40% strain and was then cut into two halves for subsequent microstructure characterization via ex-situ and quasi in-situ EBSD investigations. In order to gain insights into the evolution of texture during recrystallization, the contributions from dynamic and static recrystallization were initially separated and the origin of the non-basal orientation of recrystallization nuclei was traced back to several potential nucleation sites within the deformed matrix. Considering the significant role of double-twin band recrystallization in determining the recrystallization texture, this type of recrystallization nucleation was further investigated via quasi-in-situ EBSD on a deformed sample, annealed at 400° for different annealing times. With progressive annealing a noticeable trend was observed, in which the basal nuclei gradually diminished and eventually vanished from the annealed microstructure. In contrast, the off-basal nuclei exhibited continuous growth, ultimately becoming the dominant contributors to the recrystallization texture. The study therefore emphasizes the importance of particular nucleation sites that generate favorably oriented off-basal nuclei, which over the course of recrystallization outcompete the neighboring basal-oriented nuclei in terms of growth, and thereby dominate the recrystallization texture.

# 1 Introduction

Texture is one of the main properties governing formability in magnesium alloys. Accordingly, the study of texture formation might be the key to the design of highly formable magnesium alloys [1]. In common magnesium alloys, the hexagonal crystal structure and therefore limited activation of non-basal slip, leads to a strong basal texture rendering the material incapable of accommodating stresses out of the basal plane, i.e. along the c-axis [2, 3]. In that respect, it was found that magnesium-rare earth alloys are especially suitable to produce soft and weak textures through fine tuning of the alloy chemistry and processing parameters [4]. Binary Mg-RE alloys readily show major texture improvements characterized by weaker intensities and off-basal components in the rolling direction. New developments of ternary Mg-RE-Zn systems showed that respective alloys develop even more favorable textures with a characteristic spread of the basal poles in the transverse direction. Moreover, these alloys show a finer recrystallized grain size and a higher strength making them great candidates for sheet forming operations [5-13].

Despite extensive research efforts in that field, particular details of how new texture components nucleate during deformation or early recrystallization, and come to dominate during growth remain elusive [6, 10, 14, 15]. From segregation studies, RE solute is known to segregate to grain boundaries, and hence affect their migration behavior during recrystallization and grain growth [16-18]. This is thought to lead to the development of RE-texture components through competitive growth between basal and non-basal nuclei. The role of nucleation remains in comparison harder to elucidate due to the small scales embroiled. Nonetheless we can reasonably assume that favorable nucleation conditions for RE orientations do exist since they were seen to dominate the final recrystallization texture. Common nucleation sites reported for magnesium alloys are deformation twins, grain boundaries and triple junctions, shear or deformation bands and second phase particles [19]. The contribution of these sites to the nuclei orientation is of crucial importance for the survival of certain orientations during further texture development in the course of nuclei growth. Particularly, nucleation at $\{10\bar{1}1\} - \{10\bar{1}2\}$ double twins and shear bands was found to generate characteristic RE texture components during static recrystallization of cold rolled Mg-RE alloys [20, 21].

Recent quasi in-situ electron backscatter diffraction (EBSD) studies have provided useful insights into the origin of RE texture formation by means of tracking the evolution of microstructure during recrystallization. Guan et. al analyzed the impact of double twins [21] and shear bands on the recrystallization texture formation [20] and reported that the resultant recrystallized grains made the main contribution to the final texture development. Interestingly, the authors have also reported that concurrent events, such as precipitation can intervene with the competitive growth of basal and RE-oriented grains, thereby decreasing the substance of the desired RE components in the final texture [20]. Jiang et al. provided evidence of growth selection of RE orientations in an extruded and subsequently annealed Mg-Zn-Gd alloy depicting no significant amount of twins or shear localization [22]. Encouraged by these reports, the current study examines the influence of different nucleation sites and early nucleus growth on the texture selection during recrystallization in a Mg-Gd-Zn alloy.

## 2 Experimental

For this study, a ternary Mg-0.073at%Gd-0.165at%Zn alloy was used. Its chemical composition was measured by optical emission spectrometry (ICP/OES). An Ar/$Co_2$ gas atmosphere was applied when melting the alloy, which was cast into a pre-heated copper mold and homogenized at 460°C for 960 min. Compression tests of cylindrical samples with a diameter of 5 mm and a height of 10 mm were conducted using a ZWICK screw-driven, mechanical testing machine at temperatures between room temperature (RT) and 400°C and constant strain rate of $2 \times 10^{-3}$. The imposed deformation strains ranged between 0.1 and 0.8. The deformation conditions are summarized in Table 1.

Table 1: The investigated deformation conditions in the current work (marked by x). The sample selected for the quasi in-situ EBSD was deformed at 200°C for 40 %.

| Def. [%] \ T [°C] | RT | 200 | 400 |
|---|---|---|---|
| 10 | x | - | - |
| 20 | x | x | - |
| 40 | - | quasi in-situ EBSD | x |
| 60 | - | - | x |
| 80 | - | - | x |

Subsequent to compression, the samples were cut in half along the compression direction (Fig. 1b) and subjected to metallographic preparation using mechanical grinding and polishing with a diamond suspension up to 0.25 μm. For panoramic optical microscopy, the samples were electro-polished in an 5:3 etanol:$H_3PO_4$ solution and subsequently etched in an acitic picral solution. Multiple overlapping micrographs were taken per sample using a Leica/Leitz DMR light microscope and subsequently stitched with Image Composite Editor (Microsoft). For EBSD investigations, the respective sample was additionally electro-polished for 120 s using a Lectro-Pol 5 operating at 20 V and -20 °C in a Struers AC2 solution.

Texture measurements by X-ray diffraction (XRD) were carried out in a Bruker D8 advance diffractometer. Full pole figures and orientation distribution functions (ODFs) were calculated from 6 incomplete pole figures, $\{10\bar{1}0\}$, $\{0002\}$, $\{10\bar{1}1\}$, $\{10\bar{1}2\}$, $\{11\bar{1}0\}$, and $\{10\bar{1}3\}$, using the texture analysis toolbox MTEX [23]. EBSD measurements were conducted using a focused ion beam Helios 600i equipped with a HKL-Nordlys II EBSD detector operating at 20 kV. The resulting raw data was analyzed using the texture analysis toolbox MTEX [23]. Quasi in-situ EBSD experiments were done by interrupted heat treatments at 370°C up to 95 min in a tube furnace under protective Ar atmosphere to minimize oxidation. Between the annealing steps, the sample was shortly polished using Struers OP-U suspension. Micro-indents were used as markers for relocating the same sampled area.

## 3  Results

Fig. 1a shows the stress-strain curves after compression at different temperatures (see Table 1). Samples compressed at 400°C are characterized by a steady plastic flow at low stresses of ~ 25 MPa. On the contrary, samples compressed at RT exhibited rapid work hardening and early fracture. Both of these conditions were deemed unsuitable for further annealing experiments demanding the right amount of stored deformation energy in order to track early nucleus growth in the mapped sample. This was obtained in the sample compressed at 200°C up to 40 % strain. Despite the somewhat elevated temperature, the sample at that deformation condition still cracked due to retarded dynamic recrystallization and lack of multiple slip systems. The work softening after the peak in the respective flow curve is associated with the macroscopic cracking of the sample, evident in Fig. 2.

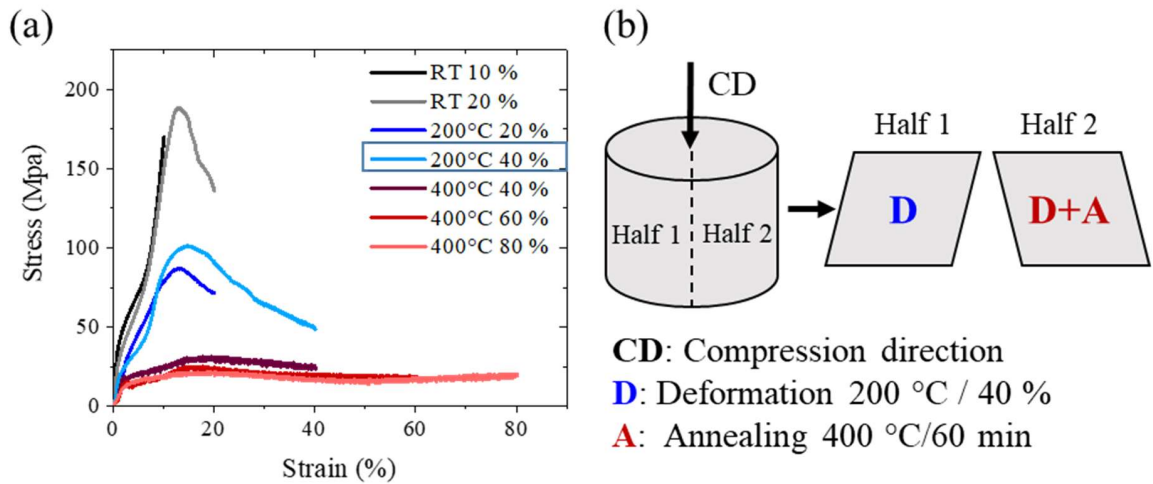

**Fig. 1:** (a) Stress-strain response during compression tests at RT, 200°C and 400°C for deformation degrees ranging from 10 to 80 %. (b) Schematic image of the cylindrical compression sample and the mirroring sample halves with their processing details.

In accordance to Fig. 1b, the chosen sample (200°C/40%) was cut in two halves along the compression direction (CD) that were characterized further by panoramic optical microscopy, EBSD and XRD, as shown in Fig. 2. The investigated microstructure in Fig. 2a revealed $\{10\bar{1}2\}$ twins that were yet unrecrystallized. This changed in the other half of the sample (half 2), which was annealed at 400°C for 60 min to initiate static recrystallization. To determine the orientation of the recrystallized grains and evaluate their nucleation mechanism, EBSD was done on both the deformed and the subsequently annealed halves of the sample. The selected EBSD areas were mirroring each other, as outlined in Fig. 2, in order to allow a correlation between the deformed and recrystallized microstructures. In addition to inverse pole figure (IPF) maps with respect to CD, the corresponding (0002), $(10\bar{1}0)$ and $(10\bar{1}2)$ pole figures are also presented in Fig 2. The provided IPF color key is applicable to all subsequent IPF maps. Hence, it is only presented once.

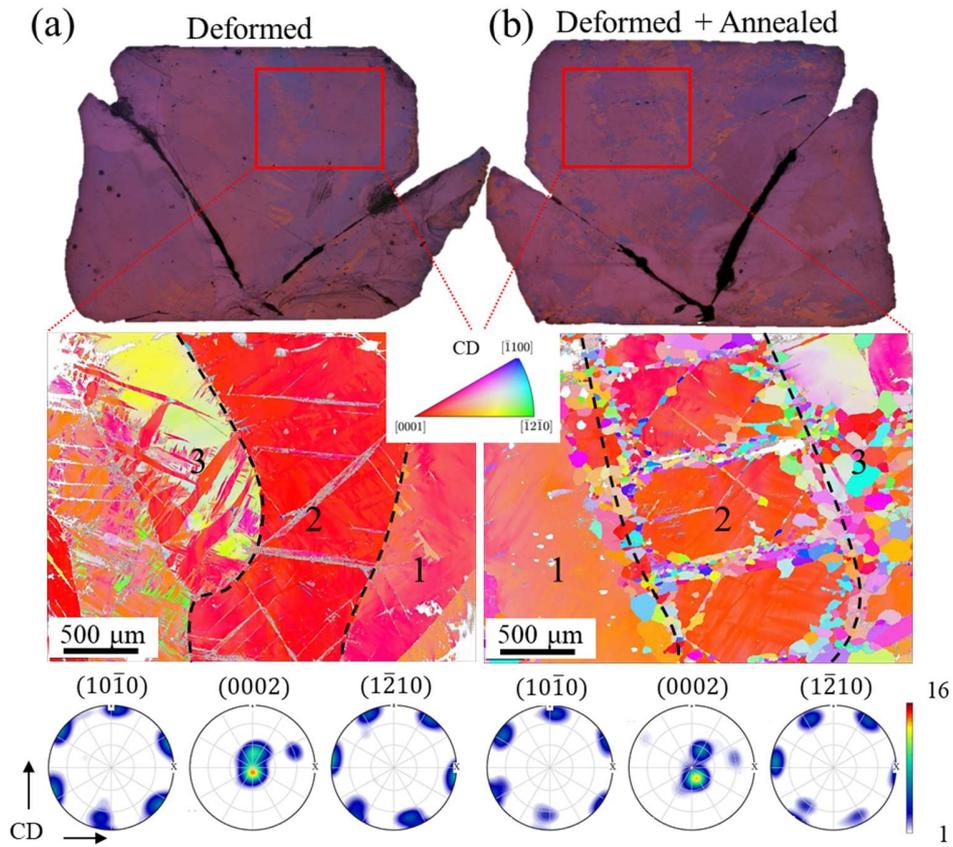

**Fig. 2:** Panoramic optical micrographs of the entire compression sample, obtained in the mid-plane parallel to the compression direction CD: (a) deformed at 200°C to 40%, (b) subsequently annealed at 400°C for 60 min. EBSD data show CD-IPF maps of the sampled areas (in the deformed and recrystallized state) and corresponding textures. The texture intensity is represented in terms of multiples of a random distribution (m.r.d.).

The left side of the CD-IPF of the deformed sample exhibited a high twin density (region 3), which subsequent to annealing underwent extensive recrystallization visible on the right side of the counterpart CD-IPF of Fig. 2b. Another feature in the deformed microstructure is a large grain with thin bands (region 2) that were liable to recrystallization nucleation and growth during annealing (Fig. 2b). As shown later in Fig. 4, those recrystallized bands are associated with double twins as they exhibited a characteristic misorientation relationship with the matrix. Further, recrystallization after annealing was seen at high angle grain boundaries separating the three big grains (regions) in the deformed microstructure. The right side of the deformed map (region 1), featuring one large matrix grain free from inner grain boundaries or obvious stress accumulations, remained free of recrystallization. The pole figures of the deformed and recrystallized microstructures do not show large qualitative differences in the overall texture since recrystallization was not advanced. It is however obvious that there was a loss in texture strength after recrystallization (from 15 to 13 m.r.d.).

To further characterize the twinning activity seen in region 3 of the deformed state, Fig. 3 presents a CD-IPF map of a higher resolution (Fig. 3b), obtained from the outlined area-of-interest in the original map (Fig. 3a). The microstructure of the selected area reveals profuse twinning of several twin variants, as well as limited dynamic recrystallization in the grain boundary region (cf. Fig. 5). The misorientation angle distribution exhibits a distinct peak around 86°±4° revealing that the observed twins were mostly $\{10\bar{1}2\}$ tension twins (TT). Additional peaks around 56°±4° and 37.5°±4° might be related to some $\{10\bar{1}1\}$ compression (CT) and compression-tension double twins (DT) (Fig. 3c). The deformation texture was found to be a characteristic basal texture, with a slight tilt of basal poles (±10°) about the compression direction (Fig. 3d).

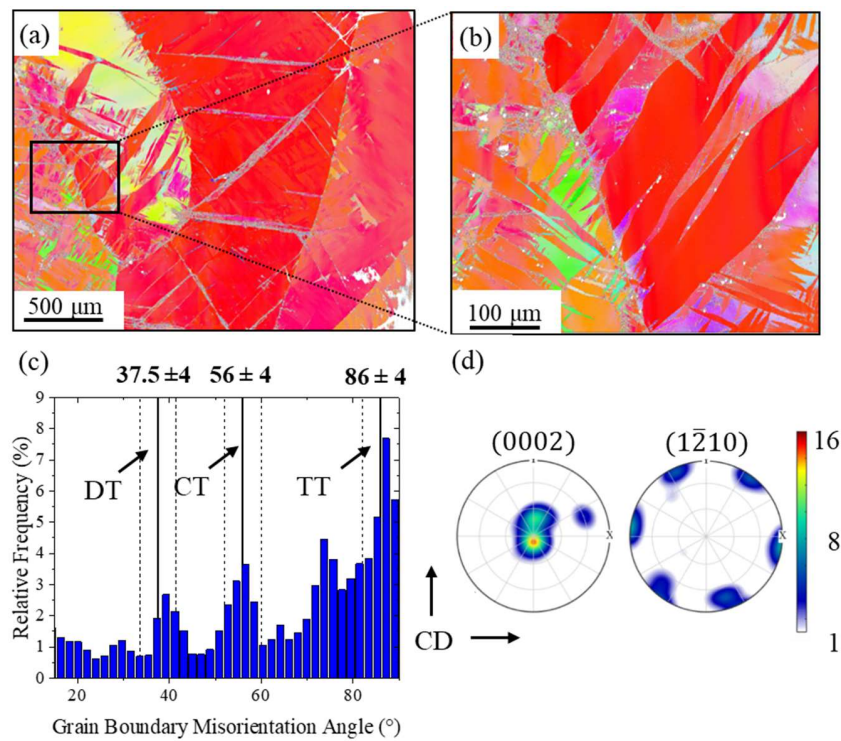

**Fig. 3:** (a) CD- IPF map after 40 % of compression at 200°C and corresponding grain boundary angle misorientation distribution (b), as well as texture from (0002), and ($1\bar{2}10$) pole figures.

Fig. 4 shows two IPF map examples showcasing the thin bands observed in the deformed microstructure (particularly in region 2, cf. Fig. 2a) and were seen to recrystallize later during annealing (Fig. 2b). Fig. 4a depicts the same deformed microstructure seen earlier in Fig. 2a, which now includes an analysis of common misorientation relationships between different twin types and their parents. As evident, the thin bands in the matrix region 2 correspond to compression-tension double-twins of the type $\{10\bar{1}1\} - \{10\bar{1}2\}$ that exhibits a characteristic misorientation of 37.5° $\langle 1\bar{2}10 \rangle$ with the parent matrix. These twins are also evident in the IPF map in Fig. 4b, which shows a different region of the microstructure after a short annealing at 370°C for 5 min. These parameters were chosen carefully to obtain evidence of the onset of recrystallization in these particular nucleation sites. The region shown in Fig. 4b is utilized later for the quasi in-situ annealing experiments presented in Fig. 8 to gain insights into the selective growth behavior exhibited by the off-basal oriented nuclei. The labelled twins TT1-3 in the matrix region 3 (cf. Fig. 2a) correspond to tension twin variants examined later in Fig. 7 in relation to recrystallization nucleation in that region.

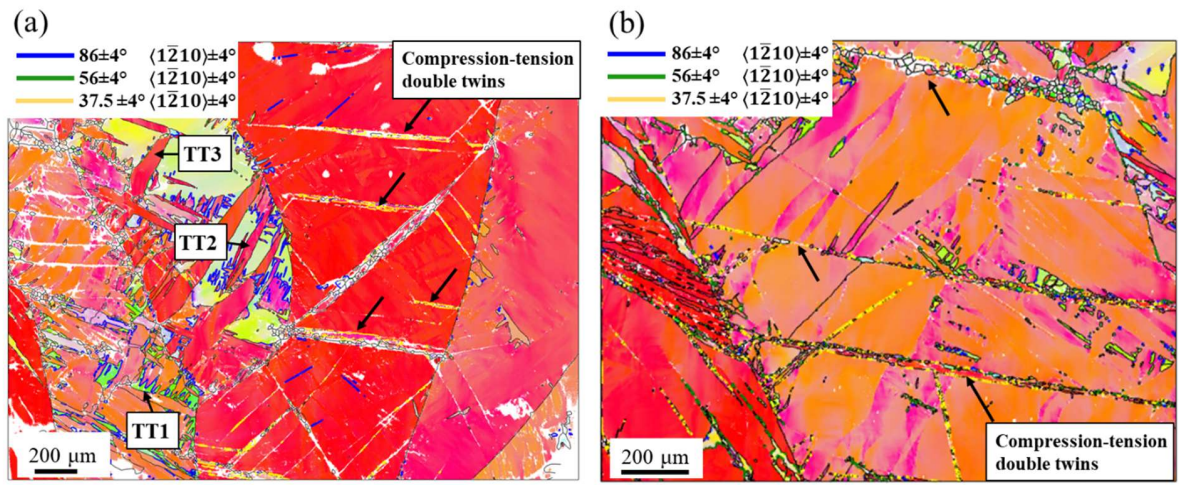

**Fig. 4**: EBSD CD-IPF maps showcasing the $\{10\bar{1}1\} - \{10\bar{1}2\}$ double twin bands in the deformed microstructure (a) and their role as a nucleation site for recrystallization in an annealed sample (370°C / 5 min) (b). TT1-3 in (a) refer to three different variants of $\{10\bar{1}2\}$ tension twins. Due to the coarse grain structure in the deformed and annealed samples, local orientation gradients within individual grains are clearly visible.

## 4  Discussion

To understand the development of texture during early stages of recrystallization, the contributions from dynamic and static recrystallization have to be separated and assigned to the different potential nucleation sites present in the microstructure (grain boundaries, tension twins and double-twin bands). The deformed area from the compression test at 200°C up to 40% strain showed limited dynamic recrystallization near the grain boundaries (GB). One example is shown in Fig 5a and b featuring a selected area of few recrystallized grains, detected using a grain orientation spread criterion (GOS <1) (Fig. 3c). The lack of sufficient number of DRX grains in the map does not allow for sound correlations to be made. However, it can be seen by the single orientations of these grains and their corresponding (0002) pole figure that some of them developed a new texture component located at 60° from the compression direction (Fig. 5e). It is also visible that the rest of the DRX grains share a similar texture component (30° from CD) with the deformed matrix (Fig. 5d and e).

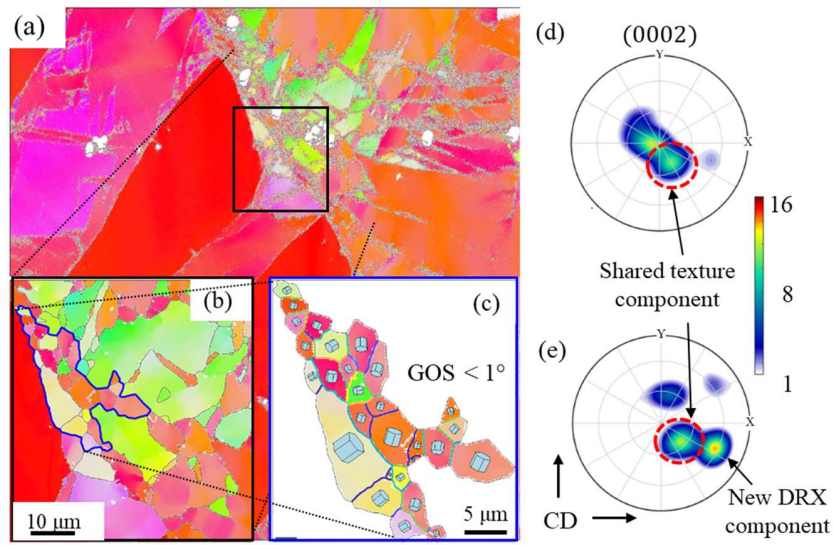

**Fig. 5:** Selected area from the CD-IPF map of the deformed sample presented in Fig. 3b featuring a dynamically recrystallized region outlined in blue (b), which is characterized by means of grain orientation spread (GOS) < 1° (c) and corresponding (0002) pole figures for the whole map (d) and the dynamically recrystallized region (e), respectively.

Retardation of dynamic recrystallization at 200°C is caused by the segregation of RE solute atoms to grain boundaries leading to solute drag [24]. This intended effect is beneficial for further static recrystallization because a sufficient amount of deformation will still be retained in the deformed microstructure. Indeed, as will be shown later, static recrystallization is crucial for the formation of soft, off-basal texture components. Fig. 6 shows the distribution of the basal poles with respect to the CD for the uniquely deformed (red) and the subsequently annealed sample half (blue), along with their corresponding (0002) pole figures, as obtained from EBSD. After deformation, the distribution of basal poles shows a close alignment with the CD (sharp basal texture) but with annealing, the spread of basal poles about the CD shifts to higher angles, leading to a much weaker texture. Similar orientation distributions with large tilt angles from the principal loading axis have been previously observed in heat treated Mg-RE-Zn alloys [5, 25], but their formation from a deformation basal texture is still unclear.

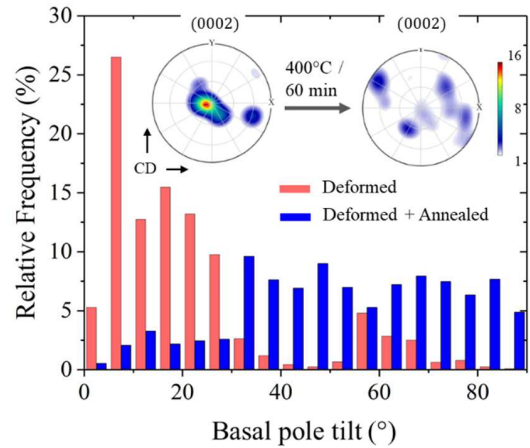

**Fig. 6:** Angle distribution between the c-axis and the compression direction obtained from EBSD data of the deformed and annealed sample. Low angles indicate close alignment of basal poles with the compression axis, i.e. a basal orientation. The insets in the figure show the corresponding (0002) pole figures of the deformed (outlined in red) and the recrystallized regions (blue) after 60 min of annealing at 400°C.

To trace back the development of the non-basal recrystallization nuclei from the deformation basal texture, the recrystallized grains from the EBSD map in Fig. 2b were carefully analyzed in connection to their original matrix site. The analysis is shown in Fig. 7, which features the isolated subset of recrystallized grains (GOS < 1°) and individual (0002) pole figures of different portions of the recrystallization microstructure. The three sets of shown pole figures in Fig. 7b (i-iii) correspond to recrystallized grains at different nucleation sites in the deformed microstructure (cf. Fig. 2a): (i) Large angle GB between matrix regions 1 and 2 (red rectangle), (ii) parallel recrystallized double-twin bands in region 2 (blue rectangle), and (iii) $\{10\bar{1}2\}$ twinned region (black rectangle). For establishing a connection with the deformation texture components, the average orientation of the parent matrix is provided in each pole figure. For (i), this information is given in terms of the average orientation of the two adjacent grains M1 and M2, for (ii) in terms of the average orientation of grain M2 and that of the double twins DT, and for (iii) in terms of the orientation of three $\{10\bar{1}2\}$ twin variants (cf. Fig. 4) labeled TT1, TT2 and TT3. The superposition of all (0002) pole figure sets produces the final (0002) pole figure for the total recrystallized IPF map, shown in Fig. 7c. The texture components in that pole figure are outlined on the basis of their origin, whether stemming from nucleation (i) at the grain boundary, (ii) within the recrystallized double-twin bands or (iii) within the heavily tensile twinned region. The latter is not only restricted to nucleation in the twin interior but also to nucleation at twin-twin interfaces and twin-GB intersections.

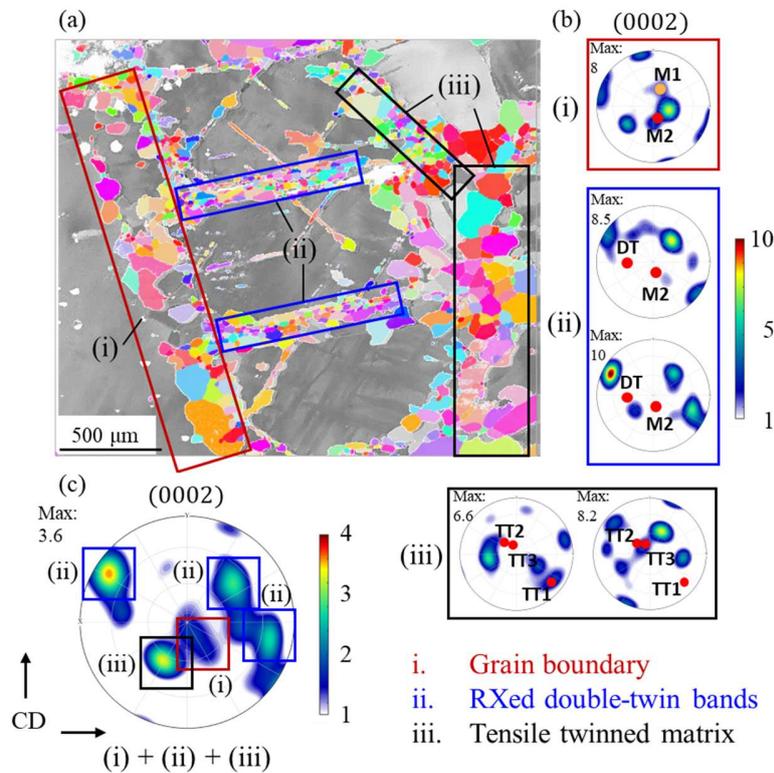

**Fig. 7:** (a) IPF subset map of the recrystallized fraction (GOS <1) reproduced from the IPF map shown in Fig. 2b. The marked regions (i) – (iii) correspond to recrystallization nucleation at a grain boundary (i), within deformation bands (ii), and within a heavily-twinned region (iii). These regions are analyzed individually by means of their (0002) pole figures (b)(i)-(iii). The texture components in the total (0002) pole figure in (c) can be traced back to their nucleation mechanism by correlating the respective orientations between (b) and (c).

From the sharpest recrystallization texture component in Fig. 7b-i, the orientation of recrystallized grains at the GB lie close to the orientation of the deformed parent matrix M1 and M2. For the tensile twinned region 3 (cf. Fig. 2), the sharpest components in the pole figures (Fig. 7b-iii) do not overlap with each other nor with the displayed orientations of the three twin variants. They show basal poles with a 30° tilt from the CD. With respect to recrystallization within the parallel double-twin bands, corresponding recrystallization nuclei revealed clear off-basal components at 60° and 90° from the CD (Fig. 7b-ii). As evident, both pole figures in Fig. 7a-ii exhibit virtually the same recrystallization texture, which seems to be quite reproducible for this nucleation site. Also visible is that this off-basal texture differs markedly from the original deformation basal texture depicted in Fig. 6. This follows the trend of texture alteration observed in Mg-RE-Zn alloys after hot-rolling and recrystallization annealing [6, 26]. The individual analysis of orientation relationships shown in Fig. 7a and b for different nucleation sites helps clarify how the final, net texture shown in Fig. 7c emerges from the competition among growing nuclei of types (i-iii) during commencing recrystallization.

Given the important contribution of double-twin band recrystallization to the final recrystallization texture, this type of recrystallization nucleation was further investigated via quasi-in-situ EBSD on a deformed sample, annealed at 400° for different annealing times between 600 s and 5700 s (95 min). The results are shown in Fig. 8 in terms of IPF maps relative to CD and corresponding (0002) pole figures. It is noted that the observation of microstructure evolution at a free surface might be affected by the reaction of grain boundaries below the surface, leading to lower average grain boundary mobility during nucleus growth. While this might have an influence on the growth rate of recrystallizing grains, it is less likely that it biases the evolution of texture in comparison to bulk recrystallization.

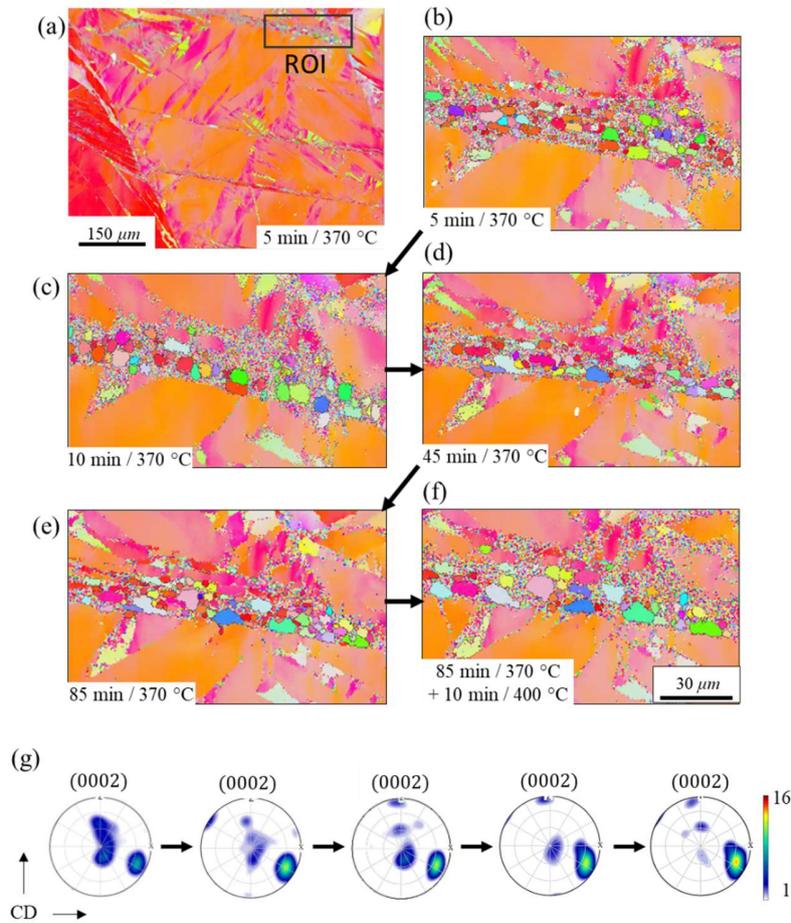

**Fig. 8:** Quasi-in-situ CD-IPF maps of a region of interest (ROI) used to track early nuclei growth in a deformation band as a function of annealing time (a-f). (g) Corresponding texture evolution of the GOS < 1° grains inside the deformation band shown in terms of (0002) pole figures.

The orientations of recrystallized grains (GOS < 1°) inside the deformation band is shown in Fig.8 g for different annealing times. The grains used for the analysis were selected manually to eliminate artifacts. The initial texture of the recrystallization nuclei after 5 min of annealing at 370°C shows two basal and off-basal orientations. With increasing annealing time, the off-basal orientation is further strengthened, while the basal orientation weakens and eventually disappear. It should be noted that the resulting main texture component after 85 min annealing time resembles the main texture components observed during ex-situ recrystallization of the double-twin bands (see Fig. 7ii). The importance of this nucleation site for the formation of favorable recrystallization texture components is therefore its ability to provide off-basal oriented nuclei capable of eclipsing basal orientations during growth. This is shown in Fig. 9 on the basis of three selected off-basal nuclei G1, G2 and G3 that grow on the expense of basal-oriented neighboring grains.

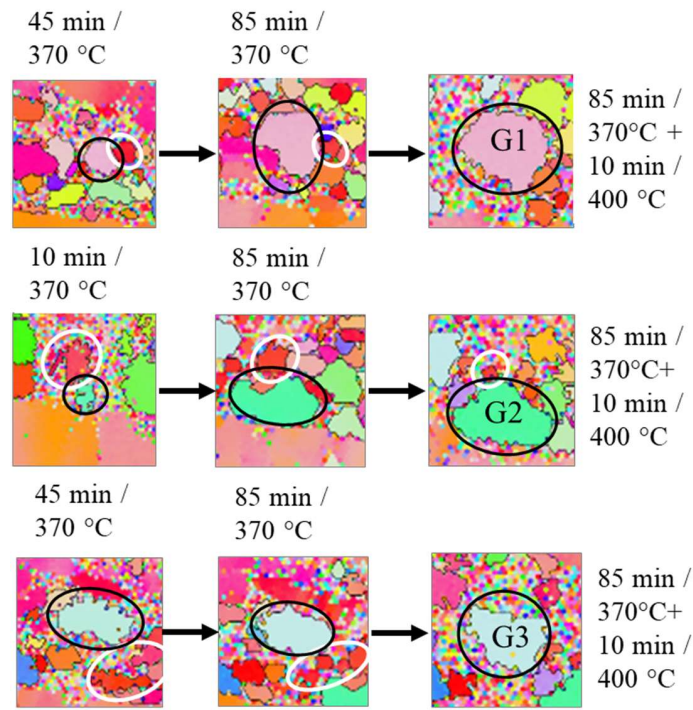

**Fig. 9:** Examples of off-basal recrystallization nuclei G1, G2 and G3 that demonstrate a selective growth behavior at the expense of shrinking, neighboring basal nuclei (white circles). This behavior explains the strengthening of the off-basal texture component shown in Fig. 7g.

Fig. 9 presents three examples highlighting the favorable growth behavior of off-basal recrystallization nuclei. These examples were reproduced from the quasi-in-situ EBSD data shown in Fig. 8. The presented cases show the off-basal nuclei G1, G2 and G3 that initially co-exist with basal-oriented neighbors of a comparable size. With progressive annealing, it can be observed how the basal nuclei (marked by white circles) gradually shrink and disappear from the annealed microstructure, whereas the off-basal nuclei grow to dominate the recrystallization texture (cf. Fig. 7 and 8). The competitive growth behavior between the basal and off-basal nuclei in the deformation bands is currently suspected to be governed by anisotropic solute segregation and resulting drag effect on the boundary migration [16, 27-29]. Solute co-segregation of both Gd and Zn to grain boundaries is to be expected, for which experimental evidence was found in previous TEM and APT studies [5, 16]. Furthermore, recent research utilizing atomistic simulations has shown that segregation to grain boundaries is of inhomogeneous nature and depends on the local atomic arrangements within the boundaries [16]. While a comprehensive understanding of anisotropic GB segregation is yet to be achieved, its effect on boundary mobility, and thus on the formation of favorable RE textures is evident. This study demonstrates the importance of particular nucleation sites in providing off-basal nuclei that prevail during incipient nucleus growth of static recrystallization at the expense of basal-oriented neighbors.

## 5 Summary and Conclusions

The current study aimed at establishing a correlation between the deformation microstructure, recrystallization nucleation, and texture selection during nucleus growth in a Mg-0.073at%Gd-0.165at%Zn alloy. For this, an as-cast sample was deformed in uniaxial compression at 200°C till 40% and then cut along the compression direction into two halves. One sample half was kept as is and the other half was annealed at 400°C for 60 min for comparative EBSD microstructure characterization of the same area. Additionally, quasi in-situ EBSD characterization was conducted on a selected area of interest featuring preferential recrystallization within a double-twin band. The following main conclusions can be drawn from the results:

1. Dynamic recrystallization in the 40% deformed sample in uniaxial compression at 200°C was restricted due to the alloy composition containing low concentrations of Gd and Zn. This was beneficial for further static recrystallization because a sufficient amount of deformation was still retained in the deformed microstructure. Texture analysis of the few dynamically RX grains revealed good potential for altering the deformation texture by forming new off-basal components.
2. Comparative EBSD microstructure characterization of the same area in the deformed and deformed + annealed samples revealed that static recrystallization after annealing at 400°C for 60 min was related to specific nucleation sites in the deformation microstructure. For clarity these sites were assigned to separate regions that included thin double-twin bands and profuse $\{10\bar{1}2\}$ twining.
3. (0002) pole figures of the deformed and recrystallized counterpart regions revealed a strong qualitative modification of the sharp basal texture developed during uniaxial deformation. The altered recrystallization texture was composed of several off-basal components that could be tracked through site-specific EBSD analysis to recrystallization within $\{10\bar{1}2\}$ – $\{10\bar{1}1\}$ double-twin bands and recrystallization within a $\{10\bar{1}2\}$ heavily-twinned region with numerous twin-parent interfaces and twin-twin intersections.
4. The influential role of double-twin recrystallization in dictating the recrystallization texture was closely examined via a combination of sequential annealing at 400°C up to 95 min and quasi-in-situ EBSD mapping. With progressive annealing, the development of texture and microstructure displayed a selective growth behavior, where the basal nuclei diminished and eventually disappeared from the microstructure, and simultaneously the off-basal nuclei grew to dominate the recrystallization texture.

**Acknowledgement.** The authors are grateful for financial support from the Deutsche Forschungsgemeinschaft (DFG), Grant No. AL1343/7-1.